\documentclass[12pt,onecolumn]{IEEEtran}

\usepackage{amsmath}
\usepackage{amssymb}  
\usepackage{amsthm}
\usepackage{float}
\usepackage{rotating} 
\usepackage{graphicx}
\usepackage{subfigure} 
\usepackage{epsfig}
\usepackage{epstopdf}
\usepackage{cite}
\usepackage{array}
\usepackage[table]{xcolor}
\usepackage{color}
\usepackage{booktabs}
\usepackage{multirow}
\renewcommand{\baselinestretch}{1.25}
\begin{document}
\title{Low Complexity Single Carrier Frequency Domain Detectors for Internet of Underwater Things (IoUT)s}
\author{\IEEEauthorblockN{Amer Aljanabi, Osama Alluhaibi, Qasim Z. Ahmed, Fahd A. Khan, Waqas-Bin-Abbas, and Pavlos I. Lazaridis}} 
\thanks{A. Aljanabi, Q. Z. Ahmed and P. Lazaridis are with School of Computing and Engineering, University of Huddersfield, United Kingdom. (Email: \{a.aljanabi,q.ahmed,p.lazaridis\}@hud.ac.uk).
O. Alluhaibi is with School of Engineering, Electrical Engineering, University of Kirkuk, Kirkuk, IRAQ. (Email: osake@uokirkuk.edu.iq)
F. A. Khan is with School of Electrical Engineering and Computer Science, National University of Sciences and Technology, Pakistan. (Email: fahd.ahmed@seecs.edu.pk)
W-B-Abbas is with Department of Electrical Engineering, FAST National University of Computer and Emerging Sciences, Pakistan. (Email: waqas.abbas@nu.edu.pk).}
\maketitle
\begin{abstract} 
This paper proposes low complexity detection for internet of underwater things (IoUT)s communication. The signal is transmitted from the source to the destination using several sensors. To simplify the computational operations at the transmitter and the sensory nodes, a single carrier frequency domain equalizer (SC-FDE) is proposed and amplify-and-forward (AF) protocols are employed. Fast Fourier transform (FFT) and use of cyclic prefix (CP) are also proposed to simplify these algorithms when compared to time-domain equalization. As precise channel data is difficult to capture in underwater communications, the adaptive implementation of FDE is proposed as a solution that can be employed when the channel experiences a fast doppler shift. The two adaptive detectors are based on the least mean-square (LMS) and recursive least square (RLS) principles. Numerical simulations show that the performance of the bit error rate (BER) performance of the proposed detectors is close to that of the ideal minimum mean square error (MMSE).	
\end{abstract}  
%\linspread{2cm}
%%%%%%%%%%%%%%%%%%%%%%%%%%%%%%%%%%%%%%%%%%%%%%%%%%%%%%%%%%%%%%%%%%%%%%%%%%
\section{Introduction}
%%%%%%%%%%%%%%%%%%%%%%%%%%%%%%%%%%%%%%%%%%%%%%%%%%%%%%%%%%%%%%%%%%%%%%%%
Underwater communications have attracted great research interest and are expected to enable a wide range of internet of underwater things (IoUT)s based applications such as pollution monitoring, oceanographic data collection, disaster prevention, tactical surveillance, underwater exploration, and port security \cite{ref-1,ref-2,ref-3,ref-4,ref-5}. The practical realization of these cooperative IoUT applications depends on several factors such as reliability, real-time long-range communication, throughput, security, etc.~\cite{ref-1,ref-2,ref-3,ref-4,ref-5,ref-5a,ref-5b,ref-5c,ref-6,ref-7,ref-8,ref-9,ref-10,Cheng-2008}, and references therein. The peculiar nature of the underwater channel makes it very difficult to satisfy the above-mentioned IoUT application requirements~\cite{ref-12,ref-13}. The underwater communication poses adverse channel conditions due to high attenuation, a large number of multipath, variable propagation delay spread, and susceptibility to Doppler shift between the source and the destination \cite{ref-5}. It is also very difficult to maintain line of sight (LOS) communication which results in significant signal attenuation~\cite{ref-12,ref-13}. Furthermore, the inter-symbol interference (ISI) span is very different in shallow and deep water~\cite{ref-14}. A possible solution to address underwater propagation channel challenges is to increase the transmit power and/or a high complexity transceiver design~\cite{ref-15,ref-16}. However, this is not a feasible option due to limited available power at underwater communicating nodes~\cite{Zhao-2017}. To address above mentioned issues, a multi-hop network with the advantages of cooperative  communication is an essential requirement for reliable communication over long distances~\cite{ref-7}. Cooperative communication allows multiple sensors to form a distributed cooperative sensor network, enabling them to achieve spatial diversity which helps to save transmission power by combating the severe signal attenuation encountered over long distances \cite{Ahmed-2014}. 

Several works have highlighted the advantages of cooperative communication in IoUTs. In \cite{ref-19}, to alter the multi-path effect at the receiver, an amplify-and-forward (AF) based cooperative transmission scheme has been proposed. Particularly, a relay node amplifies the incoming acoustic signal to provide additional multi-paths to the receiver. However, in this work, only a single relay node is considered. In \cite{ref-20}, an energy-efficient cooperative acoustic communication protocol to improve network lifetime, end-to-end delay, and reduce energy consumption has been proposed. The algorithm selects the relay and the destination nodes based on the signal-to-noise ratio (SNR) and distance among the communicating nodes. However, the proposed scheme requires SNR calculation at each link which results in high computational complexity. In \cite{ref-7}, proposed a dynamic cooperative communication scheme for underwater acoustic networks where the cooperative node transmits blocks with limited redundancy to improve transmission efficiency and to reduce end-to-end delay. In \cite{ref-21}, the authors proposed a cooperative communication-based solution where the cluster nodes are selected using a $K$-mean algorithm and cluster heads are selected using conditional probability to combat the packet loss problem and to improve the lifetime of IoUTs. In contrast to \cite{ref-7,ref-8,ref-9,ref-10}, this work considers an AF cooperative communication protocol with multiple relay nodes. Furthermore, to reduce the computational complexity, a particular node AFs the received packet based only on noise variance. 

Despite the advantages of employing cooperative communication in IoUTs, the main challenge is to resolve the number of multipath~\cite{Liang-2013,Gong-2020}. This problem is further exacerbated by a large number of relays that act as sensors in IoUTs.  At the receiver, this increases the complexity of the time-domain equalization (TDE) as additional RAKE fingers are required to resolve these multipath components \cite{Liang-2013,Zhang-2018,Boopathi-2020}. As a result, the TDE for IoUTs becomes unattractive as compared to the frequency-domain methods\cite{Wang-2012,Balevi-2016,Zhang-2008,He-2017,Tu-2018}. Frequency-domain methods employ the advantages of the Fast Fourier Transform (FFT) algorithm such as orthogonal frequency division multiplexing (OFDM) or frequency domain equalization (FDE)~\cite{Wang-2012,Balevi-2016,Zhang-2008,He-2017,Tu-2018}. 
Both single carrier OFDM and FDE techniques frequently use prefixes to formulate circular convolution. These prefixes can be classified into zero-padding and cyclic prefixes. The prefix length that directly affects the spectral efficiency, can be set based on the channel characteristics. The underwater acoustic channel (UWAC) frequently spreads tens, or even hundreds of milliseconds, and therefore to avoid the inter-block interference long prefixes are used.

In \cite{Zhang-2008}, authors presented the advantages of FDE with a single carrier (SC) for a multiple-input multiple-output (MIMO) underwater communication system. In~\cite{He-2017}, authors investigated the performance of FDE and the hybrid time-frequency-domain equalizer (HTFDE) with SC for synthetic aperture underwater acoustic communications. In~\cite{ref-10}, authors investigated the optimal relay selection and power control in an OFDM-based cooperative communication. When compared to OFDM, the SC-FDE can achieve a similar bit error rate (BER) performance, but with a lower peak-to-average power ratio (PAPR). Furthermore, FFT is carried out at the receiver side in SC-FDE, therefore, the transmitter used in IoUTs will have a very basic structure~\cite{Wang-2012}. Furthermore, as cooperative communication may involve large number of nodes, therefore a a low complexity transceiver design is an essential requirement. It can be difficult maintaining a line-of-sight (LOS) due to the movement of the platform caused by ambient disturbances or propulsion. There are therefore many potential applications involving mobile platforms, such as the one involving underwater robotics \cite{ref-6,ref-7}. It also extends network coverage by providing LOS between the source and sensors, and between sensors and destination, which results in providing multiple communication links for higher data collection rates \cite{ref-8}.

In general, underwater static source nodes (S) achieve better localization accuracy but at a higher deployment cost. They are suitable for use in early warning systems for natural disasters, such as sea-quakes, tsunamis. In addition, it is used for underwater navigation to locate dangerous rocks or shoals, especially wherein the source nodes on the seabed. However, these nodes are not as reliable for use in mobile underwater sink node schemes (R1, R2, R3 and RU). These need nodes that can be deployed quickly and are tailor-made for emergency applications. Water currents have more of a negative impact on the performance of this scheme than for the others. The final scheme to be considered is one wherein water surface vehicles are used as sensor nodes for the final destination (D). These are more practical for specific tasks, such as rapid environmental assessment, water-surface ordinance reconnaissance and the detection of potential threats. A good level of control can be achieved underwater in this model, especially when the surveillance balloons are active \cite{ref-1}. Despite this, higher electromagnetic (EM) signal attenuation rates found in the underwater environment make using acoustic signal communication links underwater the more viable option \cite{ref-2}. The yellow signals lines in Fig.1 help to illustrate this. However, the underwater acoustic link coverage range is much lower than the EM signal coverage range found in the air. The former covers approximately between one hundred and several hundred kilometres \cite{ref-3}. 

Despite the advantages of using cooperative communication, the biggest challenge is extensive multipath propagation, which confirms that the number of multipaths doubles when using a single sensor \cite{ref-9,ref-10}. This is because of the source to sensor and sensor to destination channel. Time domain equalization requires many RAKE fingers to resolve these multipaths \cite{Ahmed-2008}. Even a simple matched filter has a high computational complexity \cite{ref-12}. As a result, the time domain equalization for underwater communication becomes unattractive when compared to orthogonal frequency division multiplexing (OFDM), or frequency domain equalization (FDE) \cite{ref-13}. Receivers using frequency domains are lower in complexity than those using time domains because of their employing the Fast Fourier Transform (FFT) algorithm \cite{ref-14}. When compared to OFDM, the single carrier (SC)-FDE can achieve a similar bit error rate (BER) performance, but with a lower peak-to-average power ratio (PAPR). Furthermore, FFT is carried out at the receiver side in SC-FDE, and as a result the transmitter has a basic structure. Therefore, SC-FDE is preferred in this regard \cite{ref-14}. Evidence suggests the problems encountered when using SC-FDE for underwater amplify-and-forward  (AF)  relay channels have not been addressed satisfactorily in the literature. This knowledge gap motivated the present work.

Both single carrier OFDM and FDE techniques frequently use prefixes to formulate circular convolution. These prefixes can be classified into zero-padding and cyclic prefixes. The FDE and OFDM are periodically inserted into the transmission sequence. A block-by-block processing approach is adopted at the destination, frequently treating the channel as time invariant within a given block. A zero-padding block can be convolved to the equivalent cyclic prefix (CP) block by an overlap-add operation. The prefix length, which disturbs the spectral efficiency, could be set based on the channel characteristics. The underwater acoustic channel (UWAC) frequently spreads tens, or even hundreds of milliseconds, to avoid the inter-block interference used the long prefixes. In the meantime, the upper limit of a single block duration is defined by the rate of channel fluctuations. The channel coherence time is typically given to the order of a hundredth of a millisecond.  It is necessary for the signal block duration to be as quick as possible. This is to cause the ratio between the channel coherence time and the multipath spread associated with the spectral efficiency loss, which comes because of the use of prefixes.  It is assumed that the prefix impact on the spectral efficiency is minimal if this ratio is small \cite{ref-15}.

It is highlighted that in existing literature FDE and AF relaying have only been discussed separately. Therefore, the key novelty of this work is to analyze and evaluate the performance of combined SC-FDE and cooperative communication with AF protocol. This paper addresses the problem of reliable communication between the source and the destination in an IoUTs scenario. Particularly, the data is relayed through the intermediate sensory nodes that are placed between the source and the destination. As the main motivation is to keep the design of the transmitter and sensors simple, only a low-complexity AF cooperative diversity protocol that relies on noise variance is considered. In this work, it is assumed that the sensor node would only amplify the received signal and transmit it whereas channel estimation is only performed at the destination node. For AF networks, there exist two choices for estimating the weights, 1) the source-sensor and sensor-destination weights are estimated separately, or 2) the overall source-sensor-destination weight is estimated at the destination. The latter approach achieves higher spectral efficiency and therefore has been adopted in this work. Furthermore, to reduce the computational complexity FDE is implemented that will exploit the diagonal structure of the channel matrices. 

The main novelties of the present work are as follows:
\begin{enumerate}
\item The extension of the SC-FDE using AF relaying for IoUTs with a goal to provide a simplified and low-complexity multi-hop underwater communication architecture.
\item The proposal of adaptive implementation of FDE to combat the difficulty faced in acquiring channel knowledge in underwater communications. Particularly, two types of adaptive detectors based on the principles of least mean square (LMS) and recursive least square (RLS) are proposed in this work. Furthermore, the considered FFT structure for the received data only requires diagonal elements that reduce the computational complexity significantly as compared to the available LMS and RLS algorithms in the literature.
\end{enumerate}

The simulation results show that the SC-FDE by using AF relaying for IoUTs, which has lower complexity, wherein simplicity of transmitter and sensors is maintained. In addition the simulations results, shown that the adaptive performance of these two detectors is clsoe to the ideal minimum mean square error (MMSE) detector. 

The rest of the paper is organized as: Section 2 describes the system model for SC-FDE underwater cooperative communication systems using the AF protocol. The structure of the transmitter, sensor and receiver is discussed in detail in this section. In Section 3, the optimal and MMSE detectors are discussed. In Section 4, LMS and RLS adaptive detectors are discussed at length. The difference between these detectors and the LMS and RLS algorithms in the literature is also highlighted. The simulation results are provided in section 5, and finally conclusions are given in Section 6. 

Throughout this paper, the following notations are used: Bold upper-case letters denote matrices, while bold lower-case letters denote vectors. For arbitrary matrices,  $A, A^* , A^T , A^H$ and $A^{-1}$ denote the complex, conjugate, transpose, Hermitian and inverse of the matrix $A$, respectively. $E [\cdot]$ denotes expectation, $diag(\cdot)$ stands for diagonal matrix, $|\cdot|$  denotes the absolute value and $Tr (\cdot)$ represents the trace of a matrix.

%%%%%%%%%%%%%%%%%%%%%%%%%%%%%%%%%%%%%%%%%%%%%%%%%%%%%%%%%%%%%%%%%%
\section{System Model}
\begin{figure}
	\centering 
	\includegraphics[width=1.0\columnwidth]{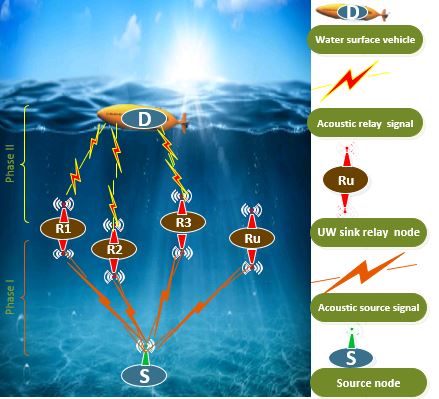}
	\caption{Data acquisition scheme of a  underwater cooperative communication system with the assistance of $U$ AF-based sensors.}
	\label{SystemModel}
\end{figure}

%%%%%%%%%%%%%%%%%%%%%%%%%%%%%%%%%%%%%%%%%%%%%%%%%%%%%%%%%%%%%%%%%%%%%5
We consider the scenario shown in Figure~\ref{SystemModel}, where a source ($S$) sends relevant data to the destination ($D$) with the help of $U$ sensors, denoted by ${R}_1,{R}_2,\dots {R}_{U}$, respectively. The sensors are assumed to possess the capability of relaying data therefore, the terms relay and sensor are used interchangeably. The source is assumed to be far from the destination and therefore no direct communication path exists between the source and the destination. The data transmission is assumed to occur over two separate phases. In phase I, the source broadcasts data to the sensors; while in phase II, the sensors amplify-and-forward the data to the destination. 
%
%%%%%%%%%%%%%%%%%%%%%%%%%%%%%%%%%%%%%%%%%%%%%%%%%%%%%%%%%%%%555
\subsection{Underwater Channel Model}
%%%%%%%%%%%%%%%%%%%%%%%%%%%%%%%%%%%%%%%%%%%%%%%%%%%%%%%%%%%%%%%555
The sound propagation speed is one of the most impactful variables on the performance of IoUTs~\cite{Liang-2013,Cheng-2008}. The propagation speed of terrestrial radio networks can by contrast reach speeds of $3 \times 10^8 $ m/s. Thus, its underwater counterpart is slower by a factor of $ 2 \times 10^5 $ m/s respectively \cite{Liang-2013}. Moreover, the complexity and spatial variability of the underwater physical environment create difficulties for IoUTs channels. 
 
The underwater acoustic channel has been shown to be composed of several distinct paths referred to as the eigenpaths \cite{Liang-2013,Cheng-2008}.  These eigenpaths are composed of a dominant and stable component as well as several smaller random components. The latter are also known as eigenarrays. Due to these characteristics a modified Saleh Valenzuela (SV) model was proposed for an underwater acoustic channel \cite{Liang-2013,Cheng-2008,Saleh-1987}. Based on this model, the channel between any two nodes is expressed as
\begin{eqnarray}~\label{eq-1}
    h^{k}_{u}(t) &=& \sum_{a=0}^{\alpha - 1} \sum_{b=0}^{\beta - 1} q^{k}_{u,b,a}\delta(t - T^{k}_{u,b,a} - \tau^{k}_{u,b,a}) \nonumber\\
&=& \sum_{l=0}^{L - 1} q^{k}_{u,l}\delta(t - \tau^{k}_{u,l}),
\end{eqnarray}
where, $k\in \{SR,RD\}$ denotes the source-relay and relay-destination hop, respectively, $\alpha$ denotes the number of eigenpaths and $\beta$ represents the number of eigenarrays for each eigenpath. For the $b$-th eigenray of the $a$-th eigenpath of the $u$-th relay in the $k$-th hop, $q^{k}_{u,b,a}$ represents the fading coefficient, $T^{k}_{u,b,a}$ is the arrival time of the $a$-th path/cluster, $\tau^{k}_{u,b,a}$ is the delay of the $b$-th ray in the $a$-th path/cluster, respectively. Equivalently, for the $l$-th eigenray of the $u$-th relay in the $k$-th hop $\tau^{k}_{u,l}$ represents the delay of the $l$-th ray, $q^{k}_{u,l}$ denotes the fading coefficient, respectively, and $L = \alpha \beta$ denotes the number of possible resolvable multipaths. 

The distribution of the cluster arrival time and the ray arrival time are modeled by a Poisson distribution and are calculated through the following~\cite{Liang-2013,Cheng-2008}
\begin{eqnarray}~\label{eq-2and3}
p(T^{k}_{u,b,a}|T^{k}_{u,b,a-1})&=&\Lambda \exp [-\Lambda(T^{k}_{u,b,a}-T^{k}_{u,b,a-1})],a>0 \\
p(\tau^{k}_{u,b,a}|\tau^{k}_{u,b-1,a})&=&\lambda \exp [-\lambda(\tau^{k}_{u,b,a}-\tau^{k}_{u,b-1,a})],  b>0
\end{eqnarray}
where, $\Lambda$ is the arrival rate of the cluster and $\lambda$ is the arrival rate of the ray within each cluster. Both arrival rates are assumed to be same for each hop. Typically, the underwater acoustic channel is quasi-static \cite{Morozs-2020,Sun-2020}. This means that the fading coefficient, $q^{k}_{u,l}$, and the delay, $\tau^{k}_{u,l}$, remain the same during one transmission burst, but then may change between bursts. The amplitude of the fading coefficient is assumed to follow independent Nakagami-$m$ distribution. The Nakagami-$m$ distribution is a generalized model using which various propagation scenarios can be modeled by the changing of the value of parameter $m$~\cite{Ahmed-2007,Simon-2005}. The probability density function (PDF) of Nakagami-$m$ distribution is given as
\begin{equation}~\label{eq-4}
    p_{|q^{k}_{u,l}|}(r) = \frac{2m_{l}^{m_l}r^{2m_l-1}} {\Gamma(m_{l})\Omega_{l}^{m_l}} \exp \left( -\frac{m_{l}}{\Omega_{l}} \right) r^2, \hspace{0.5cm} r > 0,
\end{equation}
where $\Gamma (\cdot)$ is the gamma function, $m_l$ is the fading parameter corresponding to the $l$-th multipath component and the parameter $\Omega_l=E[|h_{l}|^2] = 1$. It is assumed that the $m_l$ and $\Omega_l$ are the same for each hop~\cite{Simon-2005}. Furthermore, it is assumed that the fading channel-induced phase rotation is uniformly distributed in $[0,2 \pi]$ \cite{Ahmed-2008,Ahmed-2008a}.

%%%%%%%%%%%%%%%%%%%%%%%%%%%%%%%%%%%555
\subsection{Data Transmission}
%%%%%%%%%%%%%%%%%%%%%%%%%%%%%%%%%%%%%%%%%%5
Figure \ref{BlockDiagram} shows the block diagram of the cooperative communication for IoUTs. The communication occurring in each phase is discussed in detail below:

\begin{figure*}
\begin{center}
    \includegraphics[width=1\textwidth]{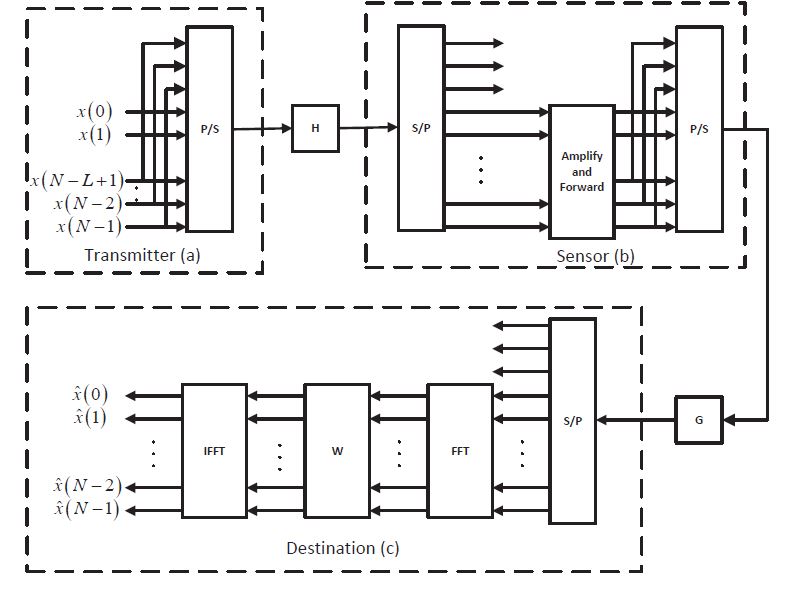}
\caption{Block diagram of a cooperative communication for IoUTs.}.
	\label{BlockDiagram}
\end{center}
\end{figure*}
\subsection*{Phase-I: Transmission from Source}
The source desires to transmit the data vector ${\pmb{x}}=[x(0),x(1),\cdots, x(N-1)]^T$ of length $N$ where $x(i)$ denotes the $i$-th data symbol chosen from a $M$-ary constellation and each symbol is assumed to be equally likely. In order to ensure circular convolution between the channel and transmitted data, so that the inter-block interference is removed through FFT-based demodulation at the receiver, a cyclic prefix (CP) of length $L$ is appended to the data vector ${\pmb{x}}$. This CP is required to have a length greater than the maximum delay profile of the channel \cite{Tu-2018}. The transmitted data vector with the CP is represented as 
\begin{equation}~\label{eq-5}
\pmb{x}_{CP}= [
\underbrace{x(N-L+1),\dots,x(N-1)}_{CP}
\underbrace{x(0),x(1),\dots,x(N-1)}_{\pmb{x}}]^T.
\end{equation}
The data vector ${\pmb{x}}_{CP}$ is converted from parallel-to-serial (P/S) prior to transmission. 
%%%%%%%%%%%%%%%%%%%%%%%%%%%%%%%%%%%%%%%%%%%%%%%%%%%%
\subsection*{ Phase-II: Transmission from the Relays to the Destination}
%%%%%%%%%%%%%%%%%%%%%%%%%%%%%%%%%%%%%%%%%%%%%%%%%%%5
The signal transmitted from the source is received at the $u$-th relay after being convolved with the respective channel, $h_u^{SR}(t)$, where $u=\{1,2,...U\}$. Each relay first converts the signal from serial-to-parallel (S/P) and discards the first $({L - 1})$ CP samples. The signal after the CP has been removed at the $u$-th relay, is expressed as
\begin{equation}~\label{eq-6}
{\pmb{y}}_{u} = {\pmb{H}}_{u}{\pmb{x}} + {\pmb{n}}_{u},
\end{equation}
where ${\pmb{n}}_u=[n_u(0),n_u(1),\cdots,n_u(N-1)]^T$ denotes the noise vector at the $u$-th relay, $n_u(i)$ is modeled as complex additive white Gaussian noise (AWGN) with mean zero and variance $\sigma^{2}_{n_{u}}/2$  per dimension and ${\pmb{H}}_u$ is a $N \times N$ channel matrix for the source to the $u$-th relay and is given in \eqref{eq:Hu}.
\begin{figure*}
	\begin{equation}
	\label{eq:Hu}
	{\pmb{H}} _u=\left[\begin{array}{cccccccc}
	h^{SR}_u(0)&0&\dots&\dots&0&h^{SR}_u(L- 1)&\dots&h^{SR}_u(1)\\
	h^{SR}_u(1)&\ddots&\ddots&\ddots&\ddots&\ddots&\ddots&\vdots\\
	\vdots&\ddots&h^{SR}_u(0)&\ddots&\ddots&\ddots&\ddots&h^{SR}_u(L-1)\\
	h^{SR}_u(L-1)&\vdots&h^{SR}_u(1)&\ddots&\ddots&\ddots&\ddots&0\\
	0&\ddots&\vdots&\ddots&\ddots&\ddots&\ddots&\vdots\\
	\vdots&\ddots&h^{SR}_u(L-1)&\vdots&\ddots&h^{SR}_u(0)&\ddots&\vdots\\
	\vdots&\ddots&\ddots&\ddots&\vdots&h^{SR}_u(1)&\ddots&0\\
	0&\dots&\dots&0&\ddots&\ddots&0&h^{SR}_u(0)
	\end{array}\right]. 
	\end{equation}
\end{figure*}
It then amplifies the remaining samples with a fixed gain $\zeta_{u}$, adds another CP and forwards the signal to the destination after P/S conversion. Without a loss of generality, the CP length at the transmitter and the relay/sensor are assumed to be same and is selected based on the channel with the maximum delay profile. As the relay node is assumed to have a low complexity, fixed gain amplifying is performed at the relay. The fixed gain $\zeta_{u}$ is expressed as 
\begin{equation}~\label{eq-8}
\zeta_{u} = \sqrt{\frac{1}{ \sigma_{h^{SR}_{u}}^{2} + \sigma_{n_{u}}^{2}}}.
\end{equation}

This fixed-gain ensures the maintenance of the average or long-term power constraint. However, it also allows the instantaneous transmit power to be much larger than the average when necessary \cite{Ahmed-2014}. The amplified received signal at the $u$-th relay after the appending the CP is given as 
\begin{equation}~\label{eq-9}
\begin{split}
&{\pmb{y}}_{u,CP}= \zeta_u \left[
\underbrace{y_u(N-L+1),\dots,y_u(N-1)}_{CP_u}\right.
\\
&\hspace{31mm}
\left. \underbrace{y_u(0),y_u(1),\dots,y_u(N-1)}_{\pmb{y}_u}\right]^T.
\end{split}
\end{equation}
%
%%%%%%%%%%%%%%%%%%%%%%%%%%%%%%%%%%%%%%%%%%55
\subsection{Receiver Structure}
%%%%%%%%%%%%%%%%%%%%%%%%%%%%%%%%%%%%%%%%%%%%%%%%%
In Fig. 2(c) the receiver converts the signal for the last time to parallel and applies the FFT transform, convolves the signal with the weight matrix $ \pmb{W}$, and applies the IFFT to detect the transmitted signals. From this figure it can be observed that the transmitter only requires a buffer to store $N$ symbols and requires a P/S convertor to carry out the transmission. Furthermore, the sensors will require S/P and P/S convertors and an amplifying and forwarding block, as well as the maximal delay profile of the channel so that an appropriate CP length can be added. Therefore, by employing an SC-FDE system, use of an inexpensive transmitter and sensors for underwater communication is possible. Each relay forwards the signal to the destination in different time slots and the signal transmitted from the $u$-th relay is received at the destination after being convolved with the respective channel, $h_u^{RD}(t)$, where $u=\{1,2,...U\}$. The destination performs S/P conversion and then discards the first $({L - 1})$ CP samples. The received signal at the destination from the $u$-th relay, after removing the CP, is given as
\begin{eqnarray}~\label{eq-10}
\pmb{r}_u &=& \pmb{G}_{u}\pmb{y}_{u} +\pmb{n} _{D,u}, \\
 &=& \underbrace{\pmb{G}_{u}\zeta_{u}\pmb{H}_{u}}_{C{u}}\pmb{x} + \underbrace{\pmb{G}_{u}\zeta_{u}\pmb{n}_{u}+\pmb{n}_{D,u}}_{\pmb{v}_{u}}\nonumber
\end{eqnarray}
where ${\pmb{n}}_{D,u}=[n_{D,u}(0),n_{D,u}(1),\cdots,n_{D,u}(N-1)]^T$ denotes the noise vector at the destination during the $u$-th relay transmission, $n_{D,u}(i)$ is modeled as complex AWGN with mean zero and variance $\sigma^{2}_{{D}}/2$  per dimension and ${\pmb{G}}_u$ is a $N \times N$ channel matrix for the $u$-th relay to destination channel and is given in \eqref{eq:Gu}.
\begin{figure*}
\begin{equation}
\label{eq:Gu}
{\pmb{G}}_u=\left[\begin{array}{cccccccc}
h^{RD}_u(0)&0&\dots&\dots&0&h^{RD}_u(L- 1)&\dots&h^{RD}_u(1)\\
h^{RD}_u(1)&\ddots&\ddots&\ddots&\ddots&\ddots&\ddots&\vdots\\
\vdots&\ddots&h^{RD}_u(0)&\ddots&\ddots&\ddots&\ddots&h^{RD}_u(L-1)\\
h^{RD}_u(L-1)&\vdots&h^{RD}_u(1)&\ddots&\ddots&\ddots&\ddots&0\\
0&\ddots&\vdots&\ddots&\ddots&\ddots&\ddots&\vdots\\
\vdots&\ddots&h^{RD}_u(L-1)&\vdots&\ddots&h^{RD}_u(0)&\ddots&\vdots\\
\vdots&\ddots&\ddots&\ddots&\vdots&h^{RD}_u(1)&\ddots&0\\
0&\dots&\dots&0&\ddots&\ddots&0&h^{RD}_u(0)
\end{array}\right].  
\end{equation}
\end{figure*} 
It can be noted that ${\pmb{C}}_{u}$ is a circulant matrix as it is a product of two circulant matrices. Acknowledging this property helps in determining the modified LMS and RLS adaptive algorithms.

The overall noise at the destination, $\pmb{v}_{u}$, can be approximated as complex Gaussian noise with zero mean and covariance matrix \cite{Fahd-2012,Ahmed-2013}
\begin{equation}~\label{eq-12}
   \Sigma v_{u} = \zeta_{u}^2 {\pmb{G}}_{u}{\pmb{G}}_{u}^H\sigma_{u}^2 + \sigma_{D}^2{\pmb{I}}_N,
\end{equation}
where $\pmb{I}_N$ denotes and identity matrix of size $N\times N$. Finally, the combined received signals from all the relays can be expressed as
\begin{eqnarray}~\label{eq-13}
        {\pmb{r}}_D &=& \sum_{u=1}^{U}{\pmb{r}}_{u}\nonumber\\
        &=&\sum_{u=1}^{U}{\pmb{C}}_{u}{\pmb{x}}+\sum_{u=1}^{U}{\pmb{v}}_{u}\nonumber\\
        &=&{\pmb{Dx}}+{\pmb{n}}.
\end{eqnarray} 
It is worth noting here that $ \textbf{D}$ will also be circulant. This is because the summation of circulant 
matrices is also itself circulant. The FFT of the received signal yield can be calculated as 
\begin{eqnarray}
     {\pmb{r}}(f) &= {\pmb{F}} {\pmb{r}}_D  = {\pmb{FDx}} + {\pmb{Fn}}\nonumber\\
     &=\underbrace{{\pmb{FDF}}^{H}}_{\Xi}{\pmb{s}}(f)+{\pmb{Fn}},
 \end{eqnarray}
where, ${\pmb{F}}$ represents the FFT and ${\pmb{s}}(f)={\pmb{F}}{\pmb{x}}$ represents transmitted signal ${\pmb{x}}$ in the frequency domain. To detect the ${\pmb{s}}(f)$, a FDE is utilized at the receiver. After obtaining the estimate of ${\pmb{s}}(f)$, an inverse FFT is carried out to obtain the estimate of the transmitted signal ${\pmb{x}}$. In the following section, the design of the FDE for the detection of $\pmb{s}(f)$ is presented.
\section{Frequency Domain Detectors}

\subsection{Optimal Detector}

The maximal likelihood (ML) detector is an optimal detector using which $\pmb{s}(f)$ can be detected as 
\begin{eqnarray}
\hat{\pmb{s}}(f) &=&\arg \min_{\pmb{s}(f)}  \|\Sigma^{-1/2}(\pmb{r}(f) - \Xi \pmb{s}(f))\|^2\\
&=& \arg\min_{\pmb{s}(f)} \left(\pmb{r}(f)- \Xi \pmb{s}(f)\right)^H\Sigma^{-1}\left(\pmb{r}(f)- \Xi \pmb{s}(f)\right),\nonumber
\end{eqnarray}
where $\Sigma = \pmb{F} (\sum_{u=1}^{U} \Sigma_{v_{u}}) \pmb{F}^H$. The process of minimization is conducted over all the possible combinations of the constellation symbols. For larger constellations, the ML Detector is considered to be intractable. Furthermore the complexity of the ML detector has been proven to be the equivalent of $O(L^{6.5})$, wherein $L$ is the number of relays~\cite{Ahmed-2013}. The complexity increases in cooperative systems which include more relays, thus rendering the ML detector impractical. 

Linear suboptimal detectors are employed to mitigate this issue. These detectors have a lower computational complexity than their ML counterparts. 
\subsection{MMSE Detector}
A linear transformation is applied to the received signal $\pmb{r}(f)$ by multiplying it by a weight matrix $\pmb{W}^H(f)$, which minimizes the mean-square error (MSE) between the transmitted symbol vector, $\pmb{s}(f)$, and the detected symbol vector, $\hat{\pmb{s}}(f)$ i.e.
\begin{equation}\label{eq:16}
\hat{\pmb{s}}(f) = \pmb{W}^H(f)\pmb{r}(f),
\end{equation}
where the optimal weight matrix $\pmb{W}(f)$ is obtained as \cite{book:Haykin}
\begin{equation}
\label{eq:WMMSEprob}
\pmb{W}(f) = \arg\min_{\pmb{W}} E[\|\pmb{s}(f) - \pmb{W}^H(f)\pmb{r}(f)\|^2],
\end{equation}
\eqref{eq:WMMSEprob} can be solved to yield the optimum weight matrix $\pmb{W}(f)$ as
\begin{equation}
\label{eq:WMMSE}
\pmb{W}(f) = \pmb{R}^{-1}(f)\pmb{\Psi}(f),
\end{equation}
where $\pmb{\Psi}(f)=E[\pmb{r}(f)\pmb{s}(f)^H]$ represents the cross-correlation between $\pmb{r}(f)$ and $\pmb{s}(f)$, while $\pmb{R}(f)=E[\pmb{r}(f)\pmb{r}(f)^H]$ is the auto-correlation of $\pmb{r}(f)$. These can be represented as 
\begin{eqnarray}
\label{eq:Psi}
&\hspace{-11mm}\pmb{\Psi}(f)= \Xi = {\pmb{FDF}}^{H},\\
\label{eq:Rb}
&\pmb{R}(f) = \Xi\Xi^H + \Sigma = \pmb{FD} \pmb{D}^H \pmb{F}^H+\Sigma.
\end{eqnarray}

\eqref{eq:WMMSE} illustrates that the MMSE detector's complexity is determined by the calculation of inverse of matrix $\pmb{R}(f)$. $\pmb{R}(f)$ is a $N \times N$ matrix, therefore the computational complexity for calculating its inverse is of $O(N^3)$ \cite{Abuzaid-2015}. However, with the introduction of CP and FFT, $\pmb{\Psi}(f)$ is now a diagonal matrix. The inversion of a diagonal matrix has a significantly lower complexity of $O(N)$, resulting the MMSE detector having the complexity level of a maximal ratio combiner (MRC) \cite{Ahmed-2014}. Furthermore, it can be noted from \eqref{eq:WMMSE}, \eqref{eq:Psi} and \eqref{eq:Rb}, that channel state information (CSI) is required to compute $\pmb{W}(f)$. Determining this accurately is difficult in IoUTs for several reasons. Firstly, the received underwater signal is made up of numerous multipath components, and each of these components has very low energy. Secondly, for AF based relaying the received channel is a cascade model, consisting of the source to the sensor and the sensor to the destination channel. This further complicates the (CSI) estimation~\cite{Gao-2008,Patel-2007}. Thirdly, the noise propagation from each cooperating relay to the destination is different, as each relay experiences different channel gain. In response to these observations, adaptive algorithms are used to establish a sub-optimal solution to the optimum weight matrix $\pmb{W}$ in \eqref{eq:WMMSE}, and thus, arrive at the optimum MMSE solution achievable through iterative computing \cite{book:Haykin}.

%%%%%%%%%%%%%%%%%%%%%%%%%%%%%%%%%%%%%%%55555
 \section {Adaptive Detection}
%%%%%%%%%%%%%%%%%%%%%%%%%%%%%%%%%%%%%%%%%%%%%%
In this section, two different approaches for constructing adaptive detectors are presented. The first is a stochastic gradient based algorithm, also termed the LMS adaptive algorithm \cite{book:Haykin,Ahmed-2008}. The second approach is based on the least square (LS) adaptive algorithm, also called the recursive least squares (RLS) algorithm \cite{book:Haykin}. The LMS algorithm is low complexity and uses stochastic gradient technique to find the suboptimal weight matrix. On the contrary, the RLS algorithm gives better estimate of the matrix, as will be shown in the simulation results, but this comes at a cost of higher processing as RLS has more calculation steps compared to the LMS algorithm. As FFT and IFFT introduces a structure to the received signal, $\pmb{F}^H \pmb{WF}$ becomes diagonal and $\pmb{W}$ becomes circulant. Thus, the complete matrix can be reconstructed using only a row or a column of $\pmb{W}$ which reduces the computations significantly. The properties of $\pmb{W}$ can be understood as following: 
\begin{equation}~\label{eq-21}
\pmb{W} = diag[w_{0},w_{1},\dots,w_{(N-1)}],
\end{equation}
where, ``diag" stands for diagonal matrix. Thus, only the $N$ diagonal elements of the matrix $\pmb{W}$ need to be computed, while the LMS and RLS of the considered literature required $(N \times N)$ elements. This further reduces the computational complexity in comparison with the algorithmic counterparts proposed in the literature.

%%%%%%%%%%%5
\subsection{LMS Algorithm}
%%%%%%%%%%%%%%%%%%%%%%%%%%%%%%

As discussed previously, using stochastic gradient technique, the LMS algorithm calculates a sub-optimal weight matrix for ${\pmb{W}}$ in \eqref{eq:WMMSE}, and ensures that the mean-square error (MSE) is similar to the MSE for the ideal MMSE detector. The steps of the algorithm are outlined below.

\begin{enumerate}

\item  Initialize the weights of the filter $\pmb{w}_0 = \pmb{0}$, when lacking any priori information.

\item Choose a suitable step-size $\mu$, such as $\textbf{\emph{0}} \leq \mu \leq \frac{2}{E||\textbf{\emph{r}}(f)||^2}$.

\item Calculate the estimated error between the transmitted and detected symbols using
\begin{equation}
{\pmb{e}}_{i}(f) = {\pmb{s}}_{i}(f) - {\pmb{w}}_{i}^{H}{\pmb{r}}_{i}(f),
\end{equation}
where ${i}$ represents the frame index. 

\item Update the weight vector as
\begin{equation}
{\pmb{w}}_{(i+1)} = {\pmb{w}}_{i} - \mu   {\pmb{r}} _{i}(f) \odot {\pmb{e}}_{i}^H(f),
\end{equation}
where $\odot$ denotes the Hadamard product. 

\end{enumerate}

%%%%%%%%%%%%%%%%%%%%%%%%%%%
\subsection{RLS Algorithm}
%%%%%%%%%%%%%%%%%%%%%%%%%%%%
The RLS adaptive algorithm is capable of significantly faster convergence speed compared to the LMS adaptive algorithm. However, this comes at the cost of higher computational complexity. It uses the detected data to improve the convergence speed as well as the detection performance. The steps of the algorithm are outlined below. 

\begin{enumerate}

\item Initialize the weights of the filter $ \pmb{w}_{0}=\pmb{0} $, when lacking any priori information. 

\item Initialize the inverse of the autocorrelation matrix $ \pmb{P}_{0}(f)=\pmb{R}^{-1}(f)=\pmb{I}$.

\item Choose a suitable forgetting factor $\lambda_{RLS}$ such that $0 \leq \lambda_{RLS} \leq 1$. 

\item Calculate the gain vector $\pmb{k}_i(f)$ as 
\begin{equation}
{\pmb{k}}_{i}(f) = \frac{\lambda_{RLS}^{-1} {\pmb{P}}_{(i-1)}(f){\pmb{r}}_{i}(f)}{1 + \lambda_{RLS}^{-1}{\pmb{r}}_{i}^{H}(f){\pmb{P}}_{(i-1)}(f){\pmb{r}}_{i}(f)},
\end{equation}
where $i$ represents the frame index.

\item Determine the estimated error between the transmitted and detected symbols as 
\begin{equation}
    {\pmb{e}}_{i}(f) = {\pmb{s}}_{i}(f) - {\pmb{w}}_{i}^{H}{\pmb{r}}_{i}(f).
\end{equation}

\item Update the weight vector using 
\begin{equation}
    {\pmb{w}}_{i} = {\pmb{w}}_{(i-1)} + {\pmb{k}}_{i}(f) \odot {\pmb{e}}_{i}^{H}(f).
\end{equation}

\item Update the inverse of the autocorrelation matrix as
\begin{equation}
{\pmb{P}}_{i}(f) = \lambda_{RLS}^{-1}{\pmb{P}}_{(i-1)}(f) - \lambda_{RLS}^{-1}{\pmb{k}}_{i}(f){\pmb{r}}_{i}^{H}(f){\pmb{P}}_{(i-1)}(f).
\end{equation}

\end{enumerate}

%%%%%%%%%%%%%%%%%%%%%%%%%%%%%%%%%%%%%%%%%%%%%%%%%%%%%%%%%%%%%%%%%%%%55
\section {{Simulation Results and Discussions}}
%%%%%%%%%%%%%%%%%%%%%%%%%%%%%%%%%%%%%%%%%%%%%%%%%%%%%%%%%%%%%%%%%%%%%%%%%%%%%5
In this section, numerical simulation results are presented to compare the performance of the proposed LMS and RLS algorithms with that of the MMSE and ML detection algorithms. In the simulations, the SV channel model using the following parameters was simulated; $1/\Lambda = 14.99$ns, $1/\lambda  = 0.476$ns, $\Gamma=0.024$ns and $\gamma = 0.12$ns. $\Gamma$ and $\gamma$ are the power decay coefficients for clusters and multipath, respectively. The number of multipaths for the source to sensor and sensor to destination were assumed to be the same, and the Nakagami-$m$ parameter was fixed to $m_l = 1.3$.

\begin{figure}
	\centering 
	\includegraphics[width=0.7\linewidth]{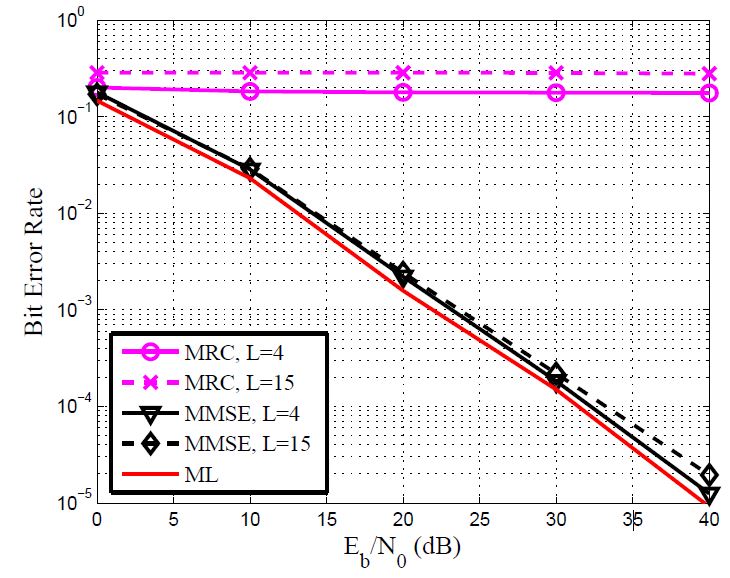}
	\caption{BER versus SNR performance of an AF underwater communication system when using a single relay with different detectors.}
	\label{fig3}
\end{figure}

For a network consisting of only one relay/sensor, Figure~\ref{fig3} shows the BER performance of the MRC, MMSE and ML detectors. Complete CSI is assumed to be available to the ML detectors. The number of multipaths considered are $L = 4$ and $L = 15$. It can be observed, that as the number of multipaths increases, the performance of the MRC detector degrades as the MRC does not have the ability to mitigate multipaths. The The BER performance of the MMSE receiver is not affected by an increased number of multipaths. It is worth noting that the complexity of MMSE and MRC is of the order $O(N)$. Its performance is also similar to that of the ML detector, which has a complexity of $O({N^{6.5}})$. Thus, it could be argued that employing SC-FDMA with the MMSE detector could resolve the multipaths issue in underwater communications.

\begin{figure}
	\centering 
	\includegraphics[width=0.7\linewidth]{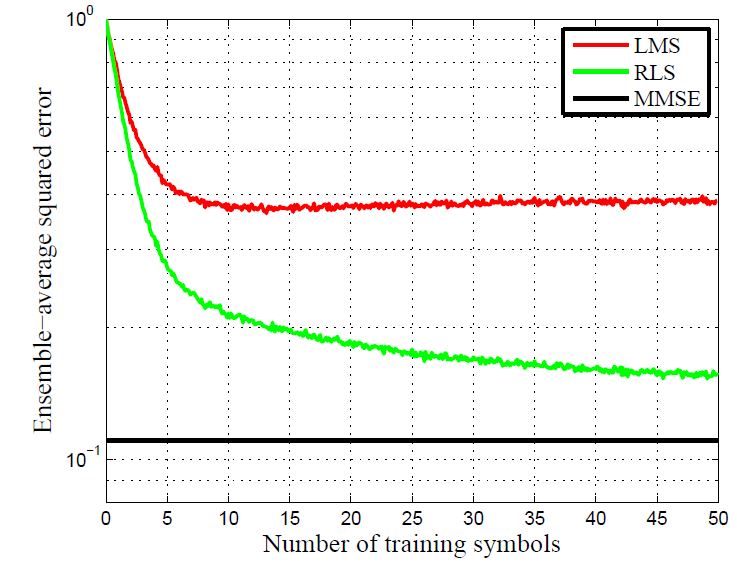}
	\caption{Convergence performance of LMS and RLS adaptive detector for underwater communication system.}
	\label{fig4}
\end{figure}

Figure~\ref{fig4} illustrates the convergence of the proposed LMS and RLS detectors for a single relay network. During the simulations, the ensemble-average squared error was obtained by averaging over 100,000 independent channel realizations at  $E_b/N_0= 5$ dB. The forgetting factor of RLS was fixed to $\lambda_{RLS} = 0.995$ while the step size of the LMS algorithm was fixed to  $\mu = 0.05$. 
It can be noted that convergence of the proposed RLS adaptive detector is superior to that of the proposed LMS based adaptive detector. Moreover, training length of 50 symbols is sufficient for algorithm convergence. 

\begin{figure}
	\centering 
	\includegraphics[width=0.7\linewidth]{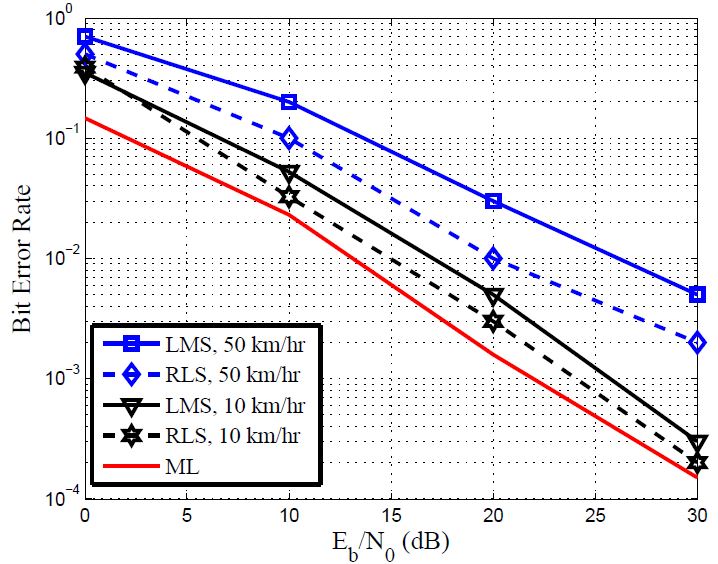}
	\caption{ BER versus SNR performance of an AF underwater communication system when using a single relay with different doppler frequency.}
	\label{fig5}
\end{figure}

Figure~\ref{fig5} compares the BER performance of both the LMS and RLS adaptive detectors when the relay is moving at different velocities. The multipath numbers were fixed to $L = 15$. 
It can be observed that, for $10$ km/hr, the RLS detector gives lower BER compared to the the LMS detector and is just 1 dB away from the ML detector which has complete CSI knowledge. This shows that the low complexity RLS detector is ideal for practical deployment. In addition, the other obvious trend which can be observed is that as the velocity increases, the BER increases.

\begin{figure}
	\centering 
	\includegraphics[width=0.7\linewidth]{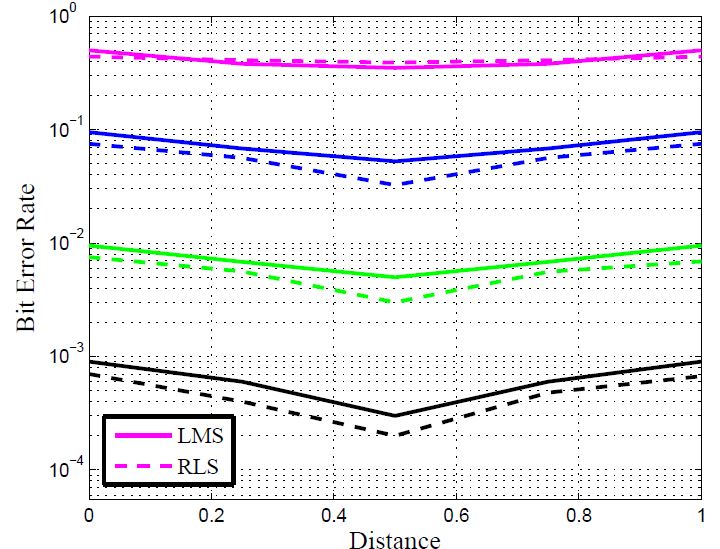}
	\caption{ BER of the AF scheme as a function of distance for underwater communication system.}
	\label{fig6}
\end{figure}

Figure~\ref{fig6} shows the effect of sensor placement when the source and sensor are transmit with the same power. The normalized doppler frequency was fixed to $0.001$. The figure reports an SNR = 0,10,20 and 30 dB respectively. The SNR decreases from bottom to top. For ease, the distance between the source and relay has been kept at $\delta$, while that between the relay and destination was $(1-\delta)$. The best position seems here to have been achieved when the relay is equidistant between the source and destination. It can also be observed that for higher SNRs, the RLS algorithm always perform better compared to the LMS detector.

\begin{figure}
	\centering 
	\includegraphics[width=0.7\linewidth]{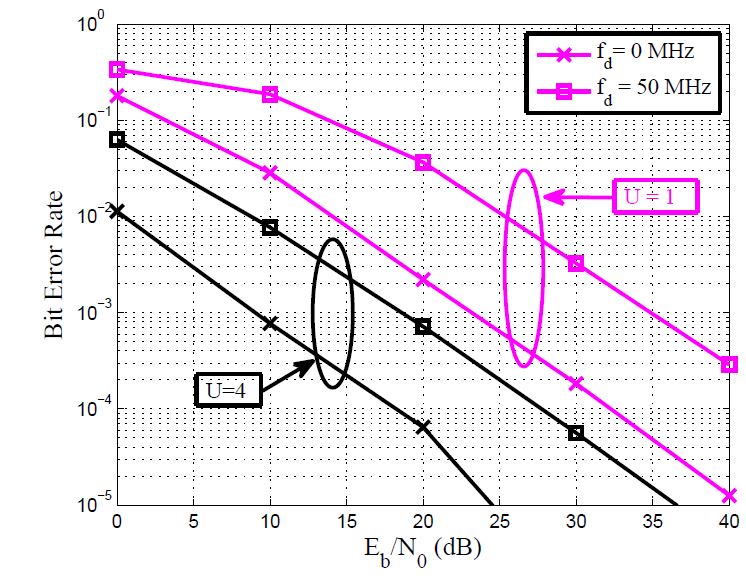}
	\caption{ BER versus SNR performance of an AF underwater communication system when using multiple relays with different doppler frequency.}
	\label{fig7}
\end{figure}

Figure~\ref{fig7} shows the impact of multiple relays on the underwater communication system. The BER is plotted for the RLS adaptive algorithm. It can be observed that as the number of relays increases, the BER performance of the detector improves. Furthermore, the RLS adaptive detector exhibits good performance even for different normalized doppler frequencies, wherein 0 MHz corresponds to the stationary relay and 50 MHZ corresponds to a movement of approximately 90 km/hr.

\section{Conclusion}

In this contribution, the use of SC-FDE detectors for underwater communication systems alongside the AF protocol has been discussed. Specifically, the ML, the ideal MMSE and two types of adaptive detectors - namely the LMS and RLS. Furthermore, the approach of measuring both the source sensor and sensor destination weights together to improve the spectral efficiency of these detectors was adopted. It can be concluded from the results herein that the considered adaptive detectors could inform numerous novel detection schemes for SC-FDE underwater cooperative communication systems. The simulation results suggest that the considered adaptive detection schemes can achieve BER performance similar to that of the ideal ML detector, which requires ideal channel knowledge. The RLS adaptive detector can achieve the best BER performance, but with a higher detection complexity than that of the LMS adaptive detector. 

%\begin{acknowledgements}
%\end{acknowledgements}



\end{document}